\documentclass{PoS}
\usepackage{epsfig}
\usepackage{graphicx}
\usepackage{amsmath}
\usepackage{multirow}
\usepackage{lscape}

\title{$N_f$=2+1 flavour equation of state}

\ShortTitle{$N_f$=2+1 flavour equation of state}

\author{
Szabolcs~Bors\'{a}nyi$^a$, Gergely~Endr\H{o}di$^b$, Zolt\'{a}n~Fodor$^{a,b}$, Antal Jakov\'ac$^a$,
S\'{a}ndor~D.~Katz$^b$, Stefan Krieg$^{a,c}$, Claudia~Ratti$^a$ and K\'{a}lm\'{a}n~K.~Szab\'o$^a$\footnote{speaker}\\
$^a$Department of Physics, University of Wuppertal, Gauss 20, D-42119,
Germany\\
$^b$Institute for Theoretical Physics, E\"otv\"os University, P\'azm\'any
1, H-1117 Budapest, Hungary\\
$^c$Center for Theoretical Physics, MIT, Cambridge, MA 02139-4307, USA
}

\abstract{
We conclude our investigation on the QCD equation of state (EoS) with
$2+1$ staggered flavors and one-link stout improvement. We extend our previous
study [JHEP 0601:089 (2006)] by choosing even finer lattices. 
These new results [for details see arXiv:1007.2580] 
support our earlier findings. Lattices with
$N_t=6,8$ and $10$ are used, and the continuum limit is approached by checking
the results at $N_t=12$. A Symanzik improved gauge and a stout-link improved
staggered fermion action is taken; the light and strange quark masses are
set to their physical values. Various observables are calculated in 
the temperature ($T$) interval of 100 to 1000~MeV.
We compare our data to the equation of state obtained by the ``hotQCD''
collaboration.
}

\FullConference{The XXVIII International Symposium on Lattice Field Theory, Lattice2010\\
		June 14-19, 2010\\
		Villasimius, Italy}

\begin{document}

{\bf Introduction}
The study of QCD thermodynamics and that of the phase diagram are receiving
increasing attention in recent years.  A transition occurs in strongly
interacting matter from a hadronic, confined system at small temperatures and
densities to a phase dominated by colored degrees of freedom at large
temperatures or densities. 
A systematic approach to determine properties of this transition is
through lattice QCD. Lattice simulations indicate that the transition at
vanishing chemical potential is merely an analytic crossover
\cite{Aoki:2006we}. This field of physics is particularly appealing because
the deconfined phase of QCD can be produced in the laboratory, in the
ultrarelativistic heavy ion collision experiments at CERN SPS, RHIC at
Brookhaven National Laboratory, ALICE at the LHC and the future FAIR at the
GSI. The
experimental results available so far show that the hot QCD matter produced
experimentally exhibits robust collective flow phenomena, which are well and
consistently described by near-ideal relativistic hydrodynamics.
These hydrodynamical models
need as an input an EoS which relates the local
thermodynamic quantities.

Most of the results on the QCD EoS have been obtained using improved
staggered fermions. This formulation does not preserve the flavor
symmetry of continuum QCD; as a consequence, the spectrum of low lying hadron
states is distorted.  Recent analyses performed by various collaborations
\cite{Huovinen:2010tv,Borsanyi:2010bp} have pointed out that
this distortion can have a dramatic impact on the thermodynamic quantities. 
To quantify this effect, one can compare the low temperature behavior of
the observables obtained on the lattice, to the predictions of the Hadron
Resonance Gas (HRG) model.

{\bf Lattice framework}
Here we present our results for thermodynamic
observables: pressure (p), trace anomaly (I=$\epsilon$-3p) and speed
of sound ($c_s$), for $n_f$=2+1 dynamical quarks. 
We improve our previous
findings \cite{Aoki:2005vt} by choosing finer lattices ($N_t=8,~10$ and a few
checkpoints at $N_t=12$). We work again with physical light and strange quark
masses: we fix them by reproducing the physical ratios $f_K/m_\pi$ and
$f_K/m_K$ and by this procedure \cite{Aoki:2006br,Aoki:2009sc,Borsanyi:2010bp} 
we get $m_s /m_{u,d}$=28.15.

\DOUBLEFIGURE{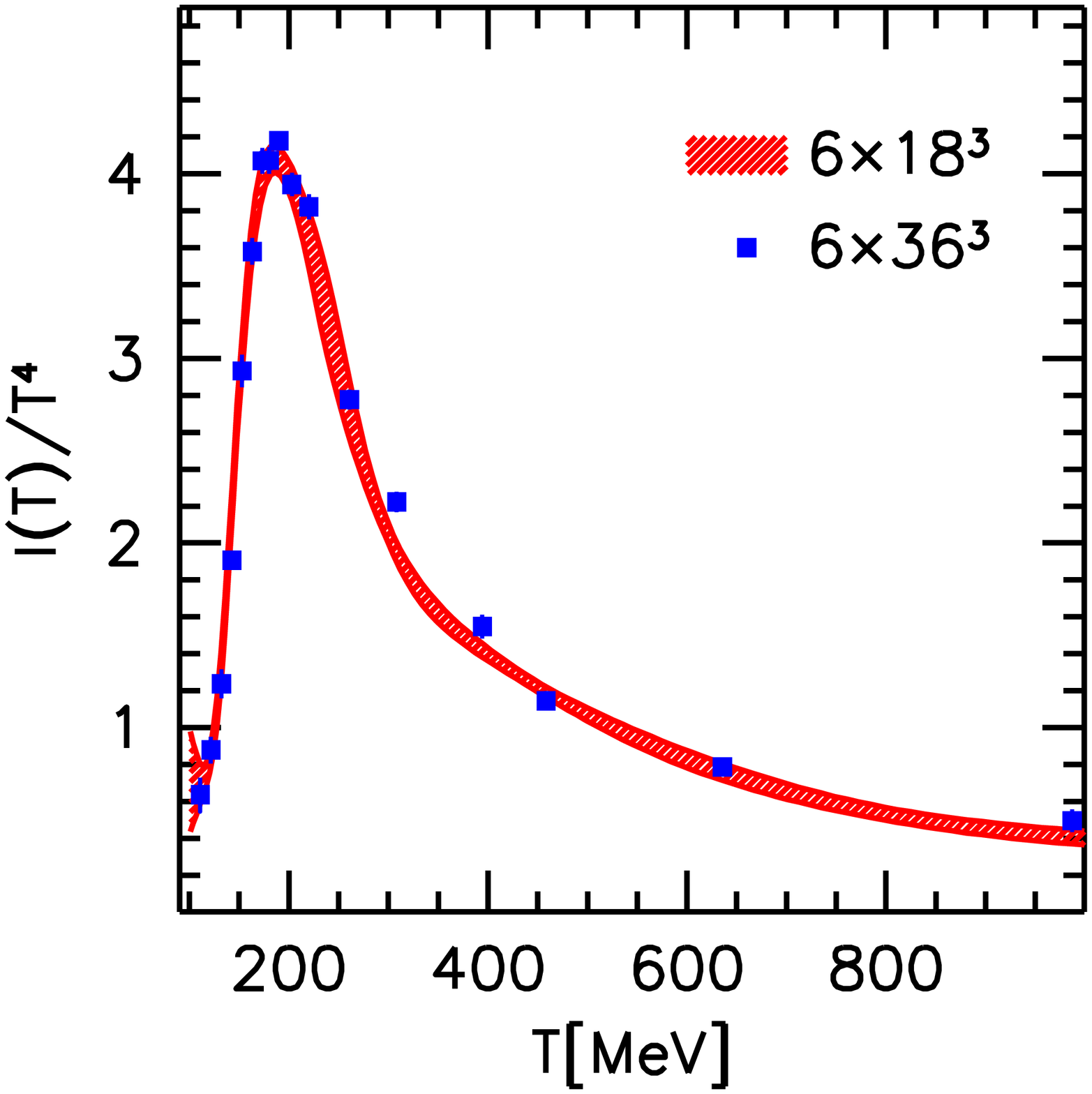,width=7.2cm}{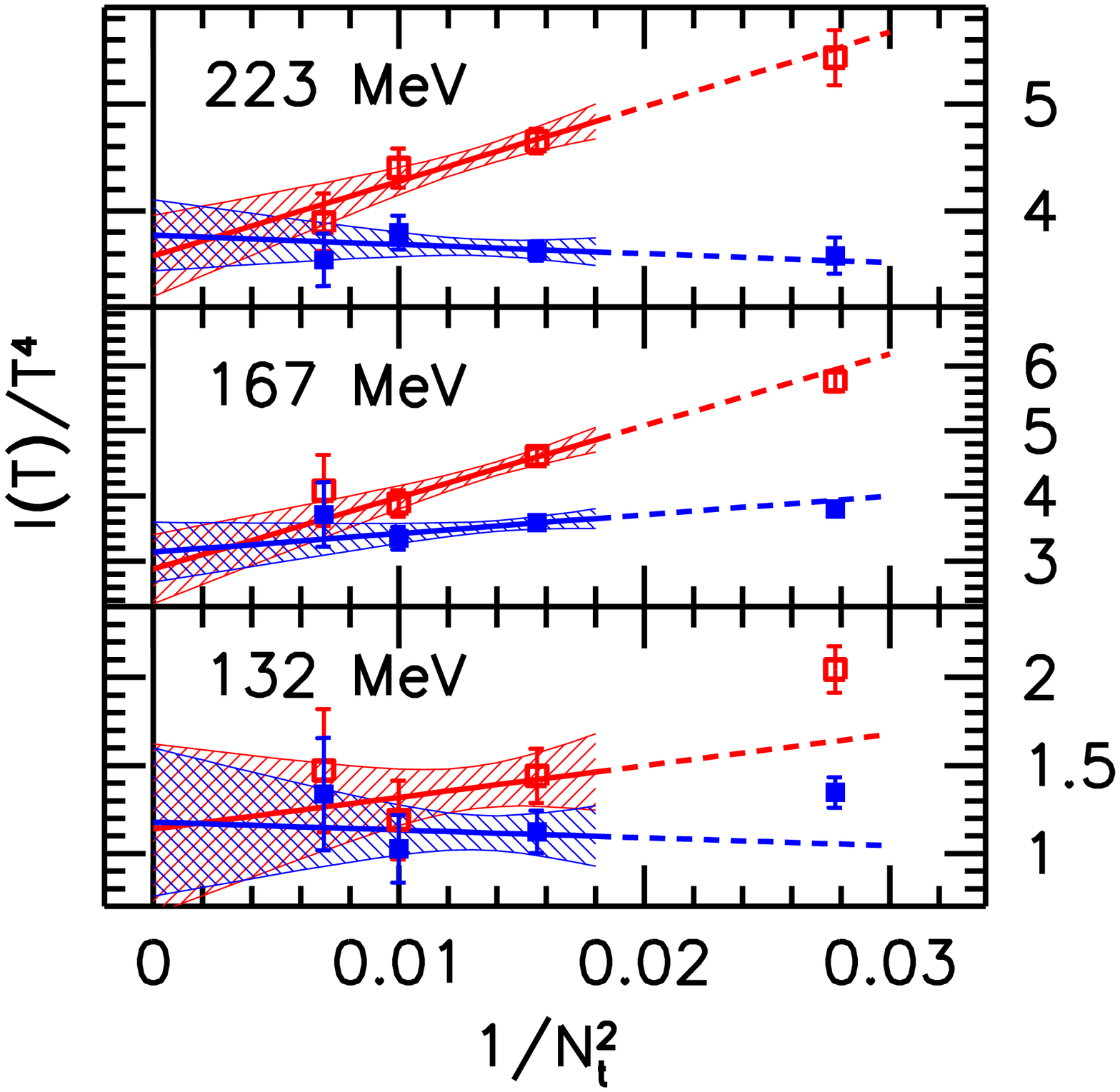,width=7.2cm}
{\label{fig:finV}
The trace anomaly on lattices with different spatial volumes:
$N_s/N_t=3$ (red band) and $N_s/N_t=6$ (blue points).
}
{\label{fig:rescale}
I=$\epsilon$-3p at three different $T$-s as a
function of $1/N_t^2$.
Filled/open symbols represent results with/without tree-level improvement.
}

We checked that there were no significant finite size effects, 
by performing two sets of simulations in boxes with a
size of 3.5~fm and 7~fm around $T_c$. Figure
\ref{fig:finV} shows the comparison between the two volumes for the
normalized trace anomaly $I/T^4$. Let us note here, that the volume independence in the
transition region is an unambiguous evidence for the crossover type of the
transition. 

\begin{figure}[p]
\epsfig{file=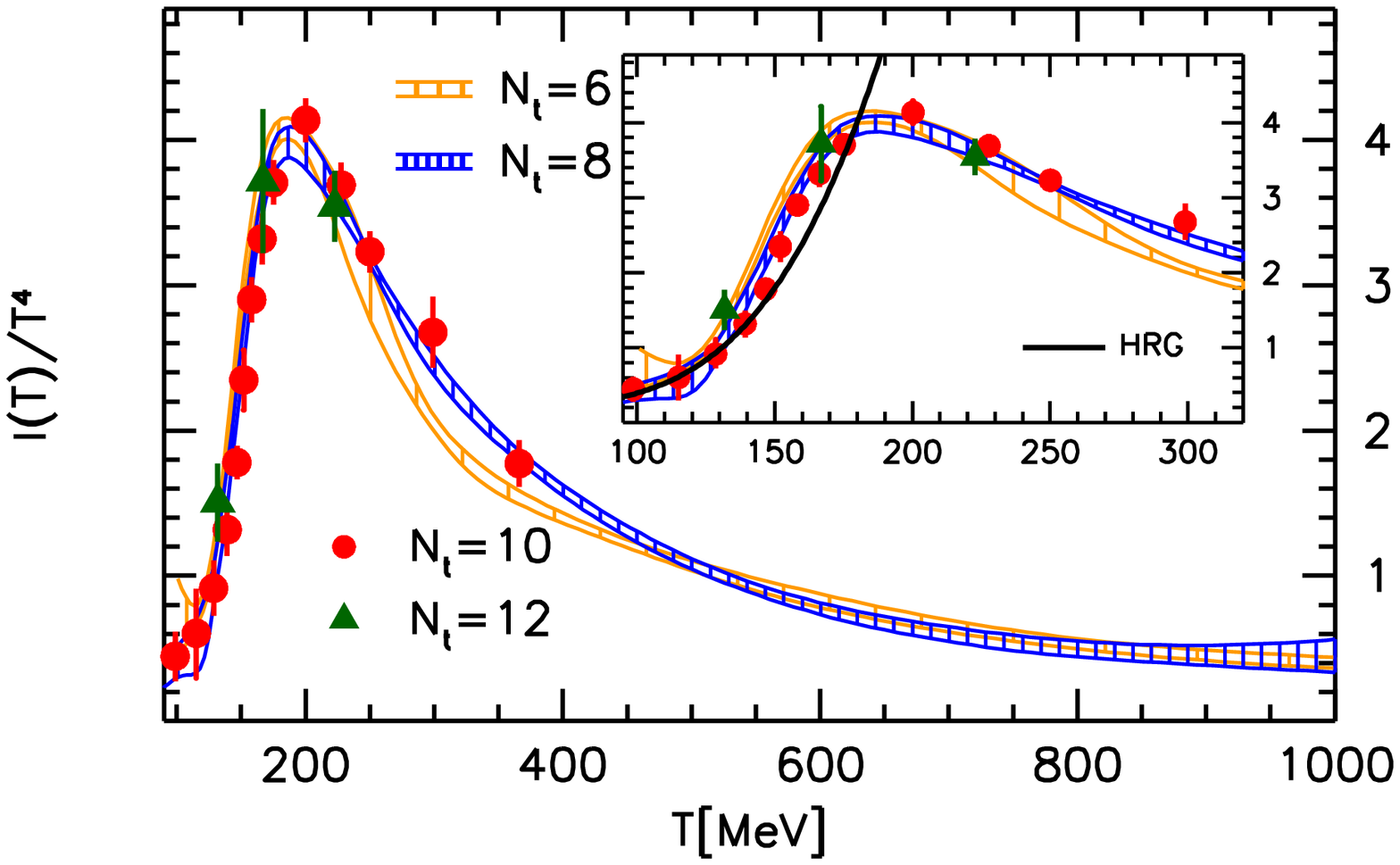,height=9cm,bb=18 360 592 718}\caption{\label{fig:eos_I}
The trace anomaly normalized by $T^4$ as a function of $T$ on $N_t=6,8
,10$ and $12$ lattices.
}
\end{figure}

To decrease lattice artefacts, we apply 
tree-level improvement for our thermodynamic
observables: we divide
the lattice results with the
appropriate improvement coefficients.
These factors can be calculated
analytically for our action and in case of the pressure we have the following
values on different $N_{ t}$'s: $N_t$=6 gives 1.517, $N_t$=8 gives 1.283, $N_t$=10
gives 1.159 and $N_t$=12 gives 1.099.
Using thermodynamical
relations one can obtain these improvement coefficients for the energy density,
trace anomaly and entropy, too. The speed of sound receives no improvement
factor at tree level. Note, that these improvement coefficients are exact only
at tree-level, thus in the infinitely high temperature, non-interacting case.
As we decrease the temperature, corrections to these improvement coefficients
appear, which have the form $1+b_2(T)/N_t^2+...$. Empirically one finds that
the $b_2(T)$ coefficient, which describes the size of lattice artefacts of the
tree-level improved quantities, is tiny not only at very high temperatures,
but throughout the deconfined phase.
Figure
\ref{fig:rescale} illustrates at three temperature values ($T=132$, $167$ and
$223$ MeV) the effectiveness of this improvement procedure. We show both the
unimproved/improved values of the trace anomaly for $N_{ t}$=6,8,10 and
$12$ as a function of 1/$N_t^2$. The lines are linear continuum extrapolations
using the three smallest $a$-s.  The $a$$\rightarrow$0
limit of both the unimproved and the improved observables converge to the same
value. The figure confirms the expectations, that lattice tree-level
improvement effectively reduces the cutoff effects. At all three $T$-s
the unimproved observables have larger cutoff effects than the improved ones.
Actually, all the three values of $b_2(T)$, which indicate the remaining cutoff
effects after tree-level improvement, differ
from zero by less than one standard deviation.

The most popular technique to determine the EoS is the integral method.
For large homogenous sytems p is proportional to the 
logarithm of the partition function. Its direct 
determination is difficult. Instead, one determines the partial 
derivatives with respect to the bare lattice parameters. 
Finally, p is rewritten as a
multidimensional integral along a path in the space of bare 
parameters. To obtain the EoS for various $m_\pi$,
we simulate for a wide range of bare parameters on the plane of $m_{u,d}$ and $\beta$
($m_s$ is fixed to its physical value).  Having
obtained this large set of data we generalize the integral method
and include all possible integration paths into the analysis 
\cite{Borsanyi:2010cj,Endrodi:2010ai}.
 
An additive divergence is present in p, which is independent of $T$.
One removes it by subtracting the same observables
measured on a lattice, with the same bare parameters but at a different $T$
value. Here we use lattices with
a large enough temporal extent, so it can be regarded as $T=0$.

{\bf Hadron Resonance Gas Model}
The Hadron Resonance Gas model has been widely used to study the hadronic
phase of QCD in comparison with lattice data. 
The low temperature phase is dominated by
pions. Goldstone's theorem implies weak interactions between pions at low
energies. As the temperature $T$ increases, heavier states
become more relevant and need to be taken into account. The HRG
model has its roots in the theorem of Ref. 
\cite{Dashen:1969ep}, which allows to calculate the microcanonical partition
function of an interacting system, in the thermodynamic limit
$V\rightarrow\infty$, to a good approximation, assuming that it is a gas of
non-interacting free hadrons and resonances \cite{Venugopalan:1992hy}.
Staggered lattice discretization has a considerable impact on the hadron spectrum. In
order to investigate these errors, we define a ``lattice HRG'' model, where in
the hadron masses lattice discretization effects are taken into account.

In order to compare the HRG model results with our additional lattice
simulations at larger-than-physical quark masses, we need the pion mass
dependence (see e.g. \cite{Durr:2008zz} for a recent lattice result)
of all hadrons and resonances included in the calculation. As we
already did in \cite{Borsanyi:2010bp}, we assume that all resonances behave as
their fundamental state hadrons as functions of the pion mass. For the
fundamental hadrons, we use the pion mass dependence from Reference
\cite{Huovinen:2009yb}. For larger-than-physical quark masses, the taste
symmetry violation at finite lattice spacing ($a$) has a milder impact on the
pressure. This also motivates to take a pion mass of about 700~MeV
 as the starting point of the
integration of the pressure at $T=100$ MeV.

{\bf Results}
As we will see different sets of data
corresponding to different $N_{ t}$ nicely agree with each other for all
observables under study: thus, we expect that discretization effects
are tiny. 

\begin{figure}[p]
\epsfig{file=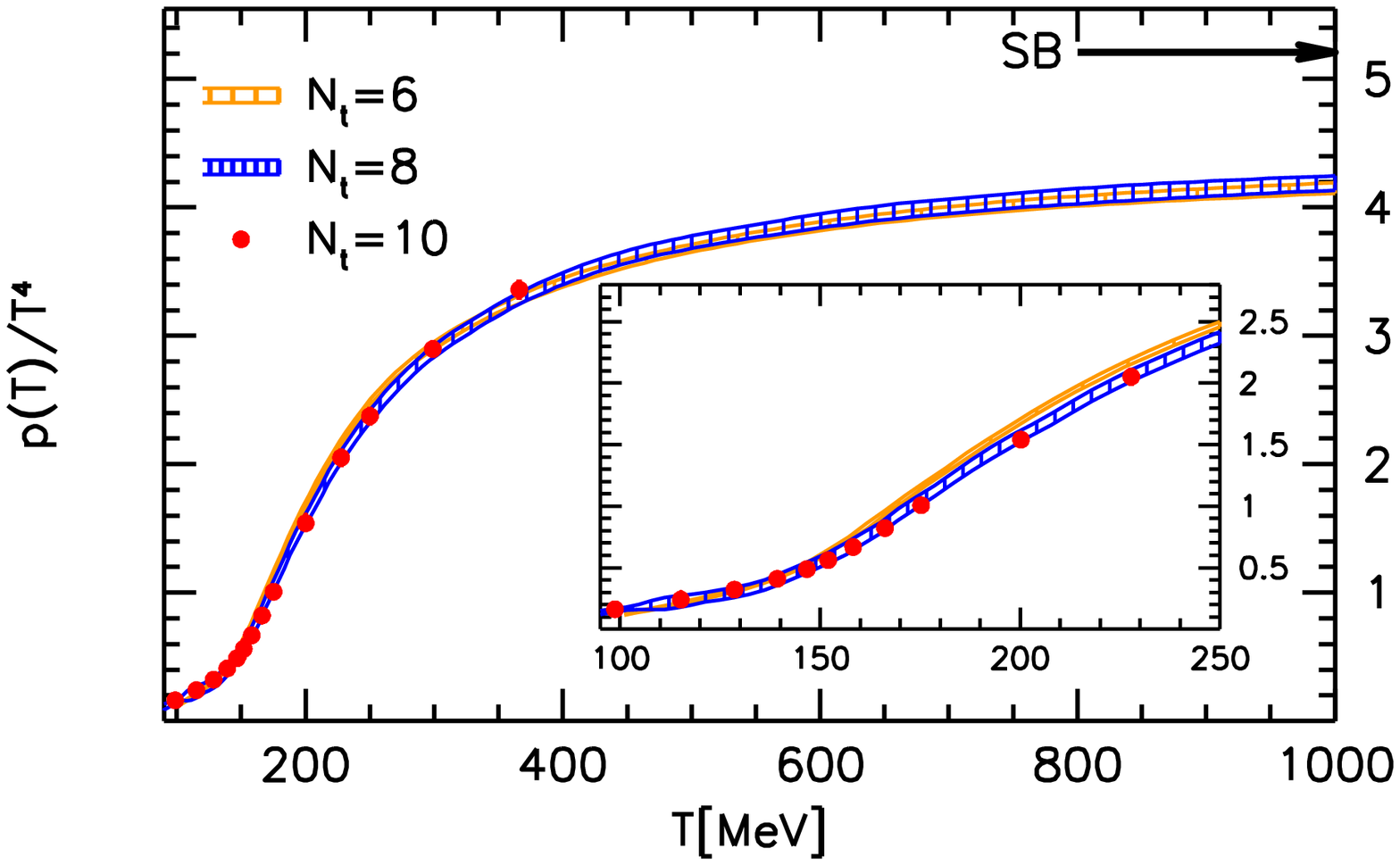,height=9cm,bb=18 360 592 718}
\caption{\label{fig:eos_p}
p(T) normalized by $T^4$
as a function of the temperature
on $N_t=6,8$ and $10$ lattices.
The Stefan-Boltzmann limit $p_{SB}(T) \approx 5.209 \cdot T^4$ is indicated. At
$T=1000$ MeV p(T) is almost 20\% below this limit.
}
\end{figure}

\begin{figure}[p]
\epsfig{file=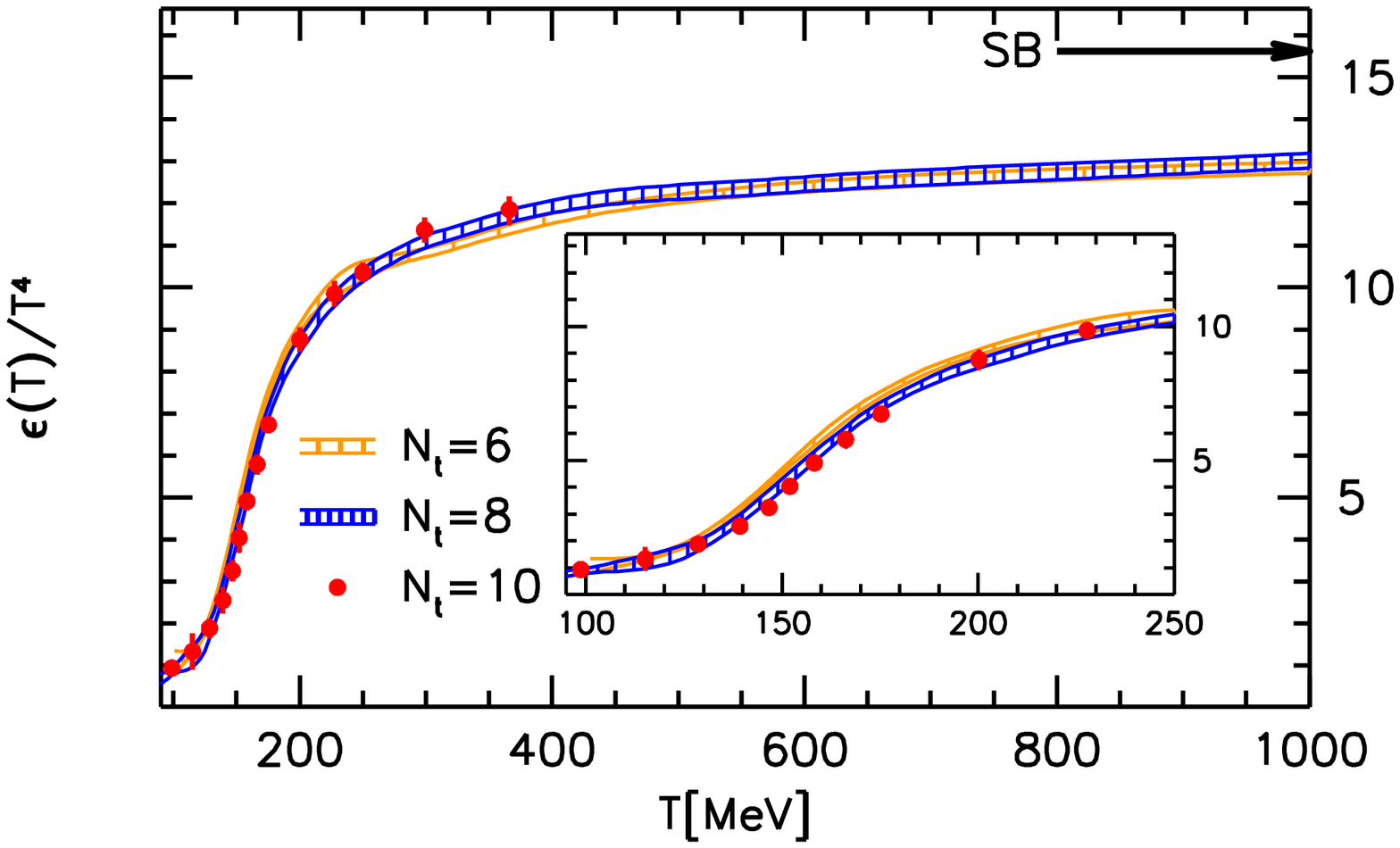,height=9cm,bb=18 360 592 718}
\caption{\label{fig:eos_e}
The energy density normalized by $T^4$
as a function of the temperature
on $N_t=6,8$ and $10$ lattices.
The Stefan-Boltzmann limit $\epsilon_{SB}= 3p_{SB}$ is indicated by an arrow.
}
\end{figure}

On Figure \ref{fig:eos_I} we show the $T$ dependence of $\epsilon$-3p for $n_f=2+1$. 
We have results at
four different "$a$"-s. Results
show essentially no dependence on "$a$", they all lie on top of
each other.  Only the coarsest $N_t=6$ lattice shows some deviation
around $\sim 300$ MeV. On the same figure, we zoom in to the
transition region. Here we also show the results from the HRG
model: a good agreement with the lattice results is found up to $T\sim 140$ MeV.
One characteristic temperature of the crossover transition
can be defined as the inflection point of the trace anomaly. This and other
characteristic features of the trace anomaly are the following:
the inflection point of $I(T)/T^4$ is 152(4)~MeV; the
maximum value of $I(T)/T^4$ is 4.1(1), whereas 
$T$ at the maximum of $I(T)/T^4$ is 191(5)~MeV.

\begin{figure}[p]
\epsfig{file=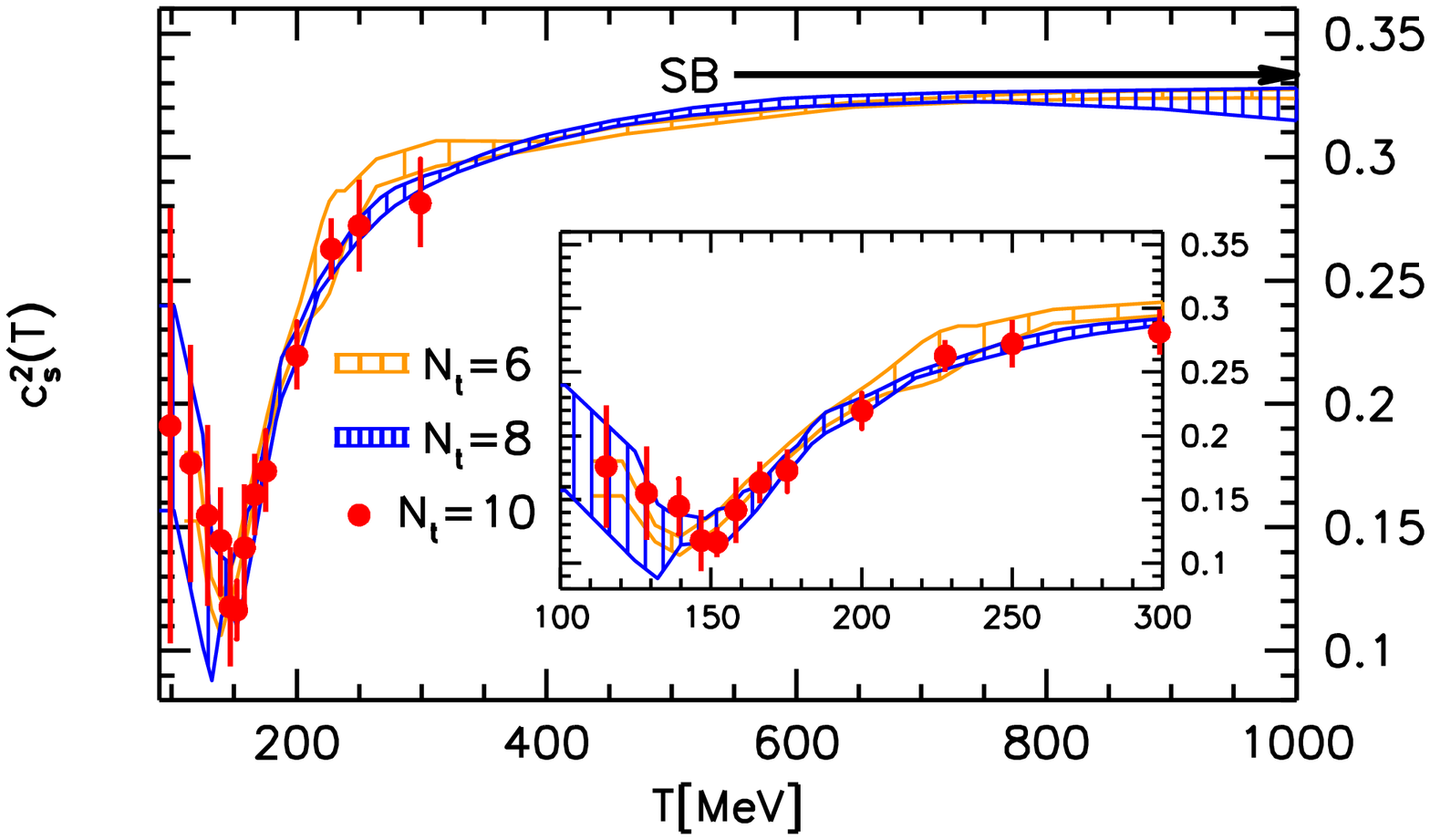,height=9cm,bb=18 360 592 718}
\caption{\label{fig:eos_cs}
The speed of sound squared
as a function of the temperature on $N_t=6,8$ and $10$ lattices. The Stefan-Boltzmann limit is $c_{s,SB}^2=1/3$
indicated by an arrow.
}
\end{figure}

On Figure \ref{fig:eos_p} we show p(T). 
We have results at three different "$a$"-s. The $N_{ t}=6$ and $N_t=8$
are in the $T$ range from 100 up to 1000 MeV. On Figure
\ref{fig:eos_e} we present the energy density.
On Figure \ref{fig:eos_cs}  $c_s^2$(T), the speed of sound is shown. 
One can also read off the characteristic
points of this curve:
the minimum value of $c_s^2(T)$ is 0.133(5); 
$T$ at the minimum of $c_s^2(T)$ is 145(5)~MeV; whereas 
$\epsilon$ at the minimum of $c_s^2(T)$ is 0.20(4)~GeV/fm$^3$.

As it was already discussed in Reference \cite{Aoki:2009sc}, there is a
disagreement between the results of current large scale thermodynamical
calculations. The main difference can be described by a $\sim$ 20-30 MeV shift
in the temperature.  This means, that the transition temperatures are
different: the temperature values that we obtain are smaller by this amount
than the values of the ``hotQCD'' collaboration. References
\cite{Huovinen:2009yb} and \cite{Borsanyi:2010bp} presented a possible
explanation for this problem: the more severe discretization artefacts of the
``asqtad'' and ``p4fat'' actions used by the ``hotQCD'' collaboration lead to larger
transition temperatures.

\begin{figure}
\begin{center}
\epsfig{file=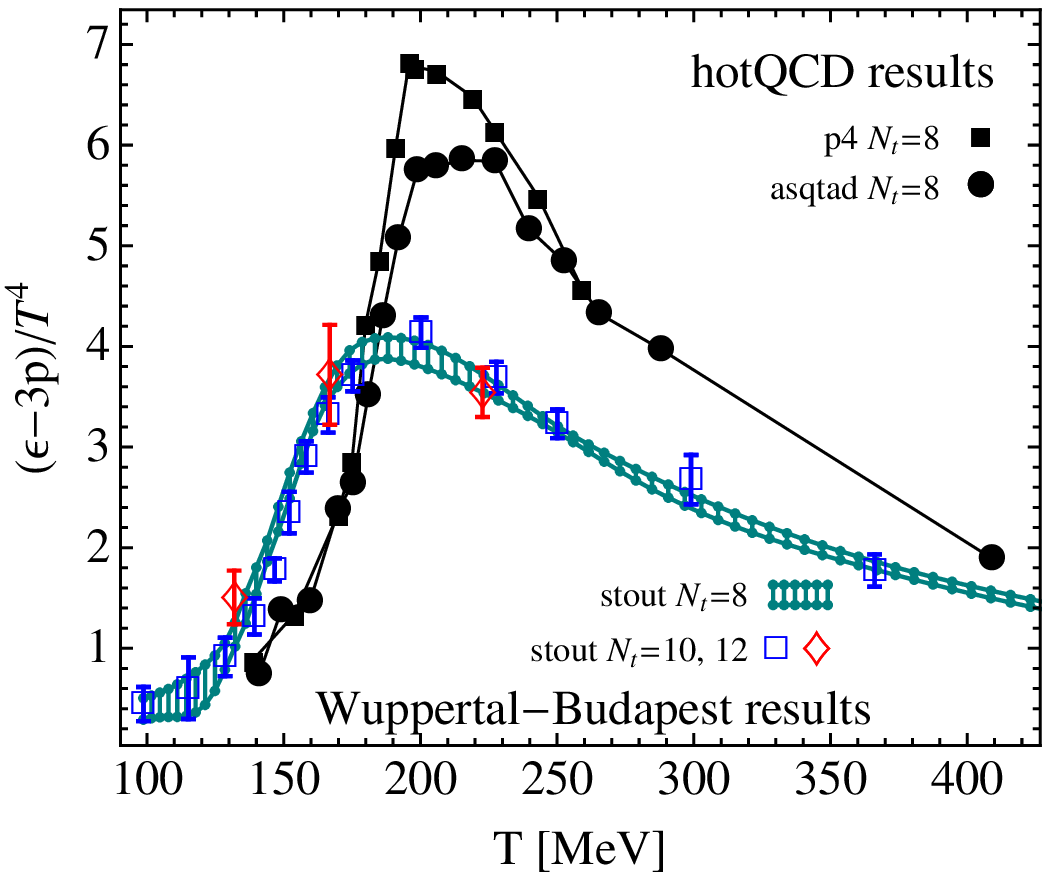,height=9cm}
\caption{\label{fig:cmp}
The normalized trace anomaly obtained in our study is compared to recent results from the ``hotQCD
'' collaboration
\cite{Bazavov:2009zn,Cheng:2009zi}.
}
\end{center}
\end{figure}

It is interesting to look for this discrepancy in the equation of state as
well. On Figure \ref{fig:cmp} we compare the trace anomaly obtained in this
study with the trace anomaly of the ``hotQCD'' collaboration.  We plot the
$N_t=8$ data using the ``p4fat'' and ``asqtad'' actions, which we took from
References \cite{Bazavov:2009zn} and \cite{Cheng:2009zi}.  As it can be clearly
seen, the upward going branch and the peak position are located at $\sim$ 20 MeV
higher temperatures in the simulations of the ``hotQCD'' group.  This is the same
phenomenon as the one, which was already reported for many other quantities in
Reference \cite{Aoki:2009sc}. We also see, that the peak height is about 50\%
larger in the ``hotQCD'' case.

{\bf Conclusions}
We determined the equation of state of QCD by means of lattice
simulations. Results for the $n_f=2+1$ flavor pressure, trace anomaly
and for the speed of sound were presented on figures. 
The results were obtained by carrying out lattice simulations at four different
"$a$"-s, at $N_t=6,8,10$ and $12$ in the $T$ range of
$T=100\dots1000$ MeV. In order to reduce the lattice artefacts we applied
tree-level improvement for all of the thermodynamical observables. We found
that there is no difference in the results at the three finest lattice
spacings. This shows that the lattice discretization errors are not significant
and the continuum limit can be reliably taken. The details of the present work can 
be found in Ref. \cite{Borsanyi:2010cj}

\end{document}